\documentclass{PoS}

\usepackage{amsmath}
\usepackage{latexsym}

\newcommand{\tr}{\mathrm{tr}}
\newcommand{\point}{\; .}

\title{About Orientifold Planar Equivalence on the Lattice}

\ShortTitle{Orientifold Planar Equivalence}

\author{\speaker{A. Patella}\\
        Scuola Normale Superiore, Pisa\\INFN, Pisa\\
        E-mail: \email{agostino.patella@sns.it}}

\abstract{The orientifold planar equivalence is the equivalence in the large-$N$ limit of the bosonic sectors of the super Yang-Mills and the QCD with a quark in the antisymmetric representation. I give a sketch of the proof of the orientifold planar equivalence in the strong-coupling and large-mass phase on the lattice. It is still matter of discussion, if its validity extends also in the continuum limit.}

\FullConference{XXIV International Symposium on Lattice Field Theory\\
                 July 23-28 2006\\
                 Tucson Arizona, US}

\begin{document}

\section{Large-N limits of QCD}

The idea behind the \textit{orientifold planar expansion} is an old one.
Because of its inherent non-perturbative features, QCD is a very hard theory
to solve. However, one can approximate QCD with some other (possibly simpler)
theory, in order to obtain analytical, if approximate, predictions. Real QCD
is an $SU(3)$ gauge theory with six quarks in the fundamental representation,
each one with a well-defined non-zero mass. Therefore, theories that
approximate real QCD can be built by slightly changing some parameters. For
instance, it is found that changing the number of colours and studying the
large-$N$ (or \textit{planar}) limit can lead to an acceptable approximation.
One can also change other parameters, like the masses of the up and
down quarks, as the chiral limit is very interesting to study.

In order to compute a large-$N$ limit, one needs to define a multicoloured
version of QCD. Usually, one chooses an $SU(N)$ gauge theory with quarks in
the fundamental representation, where the coupling constant $g^2$ is replaced
by $\lambda / N$. When the number of colours approaches infinity, this theory
becomes quenched, that is a pure gauge theory with quarks behaving as
classical external sources.

This is not the only way to generalise QCD to the case of $N$ colours.
Consider the following simple fact: the antifundamental representation of
$SU(3)$ is the same as the antisymmetric one. In fact, thanks to the
invariance of the fully skew-symmetric $\epsilon_{ijk}$ tensor, a one-to-one
map between the two representations exists.
\begin{eqnarray}
\bar{q}_i = \frac{1}{2} \epsilon_{ijk} Q^{jk} \qquad Q^{jk} = \epsilon^{ijk} \bar{q}_i \nonumber \\
\bar{q} \rightarrow U^* \bar{q} \quad \mbox{iff} \quad Q \rightarrow UQU^T
\end{eqnarray}
Thus, the multicoloured QCD can be alternatively defined as an $SU(N)$ gauge
theory with quarks in the antisymmetric representation (the coupling constant
$g^2$ must be replaced by $\lambda / N$ in this case, too). In what follows, I
will refer to this theory as \textit{orientifold QCD}. Now the question is:
what is the planar limit of orientifold QCD?

In the case of one massless flavour (but this result can be extended to the
case of more than one flavours), Armoni, Shifman and Veneziano proposed that
orientifold QCD is equivalent to super Yang-Mills in the planar limit
\cite{Armoni:2003gp,Armoni:2003fb,Armoni:2004uu}. This conjecture in known as
\textit{orientifold planar equivalence}.

In the class of strong-interacting theories, the supersymmetric ones play a
unique role. In fact, thanks to supersymmetry, some non-perturbative
quantities can be analytically computed. If the orientifold planar equivalence
holds, the corresponding quantities in QCD can be estimated up to
$1/N$-corrections. This is the case, for example, of the chiral condensate in
QCD, that can be estimated from the gluino condensate in SYM, obtaining a
value consistent with the numerical simulations \cite{Armoni:2003yv,DeGrand:2006uy}.

At the present, the validity of the orientifold planar equivalence is matter
of controversy. Armoni, Shifman and Veneziano claimed to have provided a
rigorous proof of this conjecture \cite{Armoni:2004ub}. In a recent work,
Unsal and Yaffe assert that a dynamical condition was missing in the proof:
the equivalence holds if and only if the charge conjugation symmetry is not
spontaneously broken in orientifold QCD \cite{Unsal:2006pj}.

In the future, we could get hints from numerical simulations about the
validity of the orientifold planar equivalence. Computationally, this is a
very hard problem. In fact, simulating a gauge theory with fermions in a
two-indices representation (the dimension of the representation grows like
$N^2$) has a time cost growing roughly like $N^6$. (Whilst, for a pure gauge
theory the simulation time grows roughly like $N^3$.) Deferring this problem
to future studies, it is interesting to understand what it is possible to
assert about orientifold planar equivalence between lattice-discretized
theories.

In this paper, I will give a sketch of the proof of orientifold planar
equivalence on the lattice in the phase of strong coupling and large fermionic
mass. A detailed version of this proof is available in \cite{Patella:2005vx}.

\section{Planar equivalence on the lattice}

Let us focus on the following two theories on the lattice:
\begin{description}
\item[AdjQCD.] Gauge theory with one Majorana fermion of mass $m$ in the adjoint representation.
\item[AsQCD.] Gauge theory with one Dirac fermion of mass $m$ in the antisymmetric two-indices representation.
\end{description}

In what follows, I use the Wilson discretization for the Dirac operator
\begin{equation}
D_{xy} = \delta_{xy} - \kappa \sum_\mu \left\{
( r_0-\gamma_\mu ) R[U_{\hat{\mu}}(x)] \delta_{x+\hat{\mu}, y} +
( r_0+\gamma_\mu ) R[U_{-\hat{\mu}}(x)] \delta_{x-\hat{\mu}, y} \right\} \label{diracoperator}
\end{equation}
where $\kappa$ is the hopping parameter and $R$ is the appropriate
representation which the fermions belong to. The Majorana fermion is defined
by introducing the square root of the fermionic determinant in the partition
function. Thus the (not normalized) statistical weights of the gauge
configurations are
\begin{eqnarray}
\label{weights}
\rho_{\textrm{Adj}}(U) \; \mathcal{D}U =  e^{-S_W(U)} \; \det
{D_{\textrm{Adj}}(U)}^{\frac{1}{2}} \; \mathcal{D}U \nonumber \\
\rho_{\textrm{As}}(U) \; \mathcal{D}U =  e^{-S_W(U)} \; \det D_{\textrm{As}}(U) \; \mathcal{D}U
\end{eqnarray}
for the two theories, where $S_W(U)$ is the Wilson action for the gauge field.

In the next Section, I will prove that the expectation value of a product of
Wilson loops has the same large-$N$ limit in the two theories, in the
framework of large-coupling and large-mass expansion. More precisely, if
$\left\{ \mathcal{W}_i \right\}$ are Wilson loops on the lattice, one has that
\begin{equation}
\lim_{N \rightarrow \infty} \frac{1}{N^k} \left< \mathcal{W}_1 \cdots \mathcal{W}_k \right>_\textrm{Adj} = \lim_{N \rightarrow \infty} \frac{1}{N^k} \left< \mathcal{W}_1 \cdots \mathcal{W}_k \right>_\textrm{As}
\end{equation}
where the equality holds for each term of the expansion of both the
expectation values as a power series in $\lambda^{-1}$ and $\kappa$.

\section{Planar limit in the strong-coupling and large-mass phase}

As a first step, let us perform the hopping expansion of the fermionic
effective action (for the details, see \cite{Rothe}):
\begin{equation}
S_F = - \log \det D_R^{N_f} = N_f \sum_{\alpha \in \mathcal{C}} \kappa^{L(\alpha)} c(\alpha) \mathcal{W}_R(\alpha)
\end{equation}
where $\mathcal{C}$ is the set of all the closed paths linking nearest
neighbours on the lattice, $L(\alpha)$ is the length of the path $\alpha$,
$\mathcal{W}_R(\alpha)$ is the Wilson loop along the path $\alpha$ in the
representation $R$, $c(\alpha)$ is a representation-independent coefficient and
$N_f$ is $1$ for AsQCD or $1/2$ for AdjQCD.

Using standard relations, Wilson loops in the adjoint and antisymmetric
representations can be written in terms of Wilson loops in the
(anti-)fundamental representations:
\begin{eqnarray}
&& \frac{1}{2} \tr \, \textbf{Adj}(U) = \frac{1}{2} \left\{ \left| \tr U \right|^2 -1 \right\} \simeq  \frac{1}{2} \left| \tr U \right|^2 \nonumber \\
&& \tr \, \textbf{As}(U) = \frac{1}{2} \left\{ \left( \tr U \right)^2 - \tr U^2 \right\} \simeq \frac{1}{2}  \left( \tr U \right)^2 \point
\end{eqnarray}
Here, the approximations are valid in the planar limit.

Putting all together, the actions of the two theories can be written as:
\begin{eqnarray}
S_{\textrm{Adj}}(U) \simeq \frac{2N^2}{\lambda} \sum_p \left( 1- \frac{1}{N} \textrm{Re} \, \tr \, U_p \right) + \frac{1}{2} \sum_{\alpha \in \mathcal{C}} \kappa^{L(\alpha)} c(\alpha) \left| \tr \mathcal{W}(\alpha) \right|^2 \nonumber \\
S_{\textrm{As}}(U) \simeq \frac{2N^2}{\lambda} \sum_p \left( 1- \frac{1}{N} \textrm{Re} \, \tr \, U_p \right) + \frac{1}{2} \sum_{\alpha \in \mathcal{C}} \kappa^{L(\alpha)} c(\alpha) \left( \tr \mathcal{W}(\alpha) \right)^2
\end{eqnarray}
where $\mathcal{W}(\alpha)$ is the Wilson loop along $\alpha$ in the fundamental
representation. It is clear that the orientifold planar equivalence is based
on the possibility of reversing the orientation of one of the two Wilson loops
coming from the fermionic effective action. The full strong-coupling and large-
mass expansion of the statistical weights in (\ref{weights}) is obtained by
expanding the exponential $e^{-S}$ as a power series.

In order to compute the partition function, one can replace the exponential in
the integral with its power series, and fully expand each of the terms. The
partition function is finally obtained as a sum of graphs.

Each graph is an integral over all the link variables of a product of some
plaquettes (times $2N/\lambda$) coming from the Wilson action and some couples
of Wilson loops (times $\kappa^{L(\alpha)}c(\alpha)/2$) coming from the fermionic
effective action. It can be shown \cite{collins:haar} that at leading order in
$1/N$ the effect of the integration over the gauge group is to perform
Wick-contractions between couples of $U$ and $U^\dagger$:
\begin{equation}
U_{ij}U^*_{kl} \rightarrow \frac{1}{N} \delta_{ik}\delta_{jl} \point
\end{equation}

As a consequence, we can use all the machinery developed for the usual
perturbation theory. For example, the expectation value of a Wilson loop is
given by the sum of all the ``connected graphs'' (times the appropriate
combinatorial factor) with the insertion of the Wilson loop. The ``connected
graphs'' are those which cannot be written as a product of two other graphs of
the theory.

\begin{figure}
\includegraphics[width=.6\textwidth]{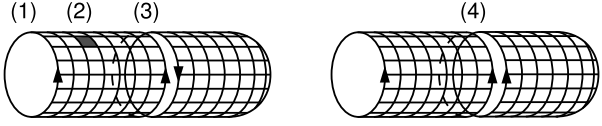}
\caption{\label{legend} On the left, a graph of the expectation value of a WIlson loop for AdjQCD. On the right, a similar graph for AsQCD. (1) Inserted Wilson loop. (2) Plaquette. (3) $ \left( \tr U \right)^2 $ from a Wilson loop in the antisymmetric representation (from the fermionic determinant). (4) $ \left| \tr U \right|^2 $ from a Wilson loop in the adjoint representation (from the fermionic determinant).}
\end{figure}

A graph can be represented as a possibly disconnected surface bounded by
Wilson loops and tiled by plaquettes, that are sewn by Wick-contractions (see
figure \ref{legend}).

As in the usual 't Hooft expansion \cite{'thooft:largeN}, graphs are
proportional to $ N^\chi $, where $ \chi $ is the Euler characteristic of the
surface:
\begin{equation}
\chi = 2C - 2H - B
\end{equation}
where $C$ is the number of connected components, $H$ is the number of handles
and $B$ is the number of boundaries.

\begin{figure}
\includegraphics[width=.8\textwidth]{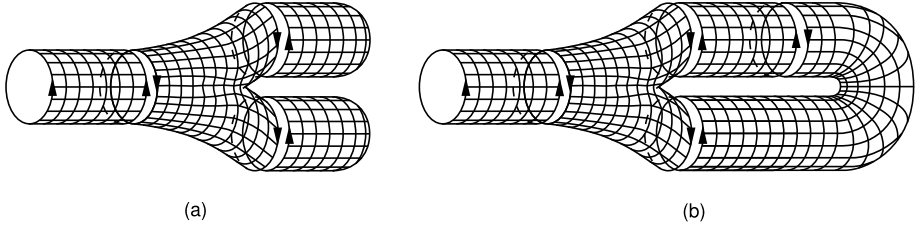}
\caption{\label{loop} Both graphs are related to AdjQCD. The graph in (a) is a planar one; its Euler characteristic is $1$. The graph in (b) is a subleading one; it contains a loop and its Euler characteristic is $-1$.}
\end{figure}

Since graphs may be represented by disconnected surfaces, it may seem that the
Euler characteristic indefinitely rises by adding connected components. This
is not the case. It can be shown that one can add connected components without
introducing subgraphs only by also adding boundaries, in such a way that the
leading power of $N$ is not modified by the presence of the fermions. In the
planar limit, only graphs with the highest possible Euler characteristic
survive. These are graphs (as in figure \ref{loop}a) without handles and
without loops (like those in figure \ref{loop}b).

\begin{figure}
\includegraphics[width=.8\textwidth]{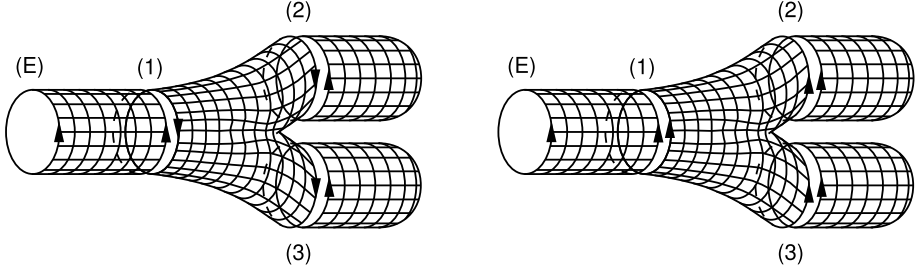}
\caption{\label{equivalence} On the left, a graph of the expectation value of a WIlson loop for AdjQCD. On the right, a similar graph for AsQCD. The latter can be obtained from the former, simply by reversing all the directions of the link variables in the central connected component. The labels in the graphs refer to the Wilson loops.}
\end{figure}

At this point, we have developed all the tools we need to prove the
orientifold planar equivalence. Let us see it with an example. Consider the
first graph in figure \ref{equivalence}. Apart from the combinatorial factor,
its value is:
\begin{equation}
\mathcal{G}_{\textrm{Adj}} = \frac{1}{8} c_1 c_2 c_3 \left< \tr U^{(E)} \, \mathcal{W}_{\textrm{Adj}}^{(1)} \, \mathcal{W}_{\textrm{Adj}}^{(2)\dagger} \, \mathcal{W}_{\textrm{Adj}}^{(3)\dagger} \, \mathcal{P}^{(E,1)} \, \mathcal{P}^{(1,2,3)} \, \mathcal{P}^{(2)} \, \mathcal{P}^{(3)} \right>_c
\end{equation}
where $\left< \cdot \right>$ represents the integration with respect to the
Haar measure. In the planar limit,
\begin{eqnarray}
\mathcal{G}_{\textrm{Adj}} \simeq && \frac{1}{8} c_1 c_2 c_3 \left< \tr U^{(E)} \, \mathcal{P}^{(E,1)} \, \tr U^{(1)} \right> \; \nonumber\\
&& \left< \tr U^{(1)\dagger} \, \tr U^{(2)\dagger} \, \tr U^{(3)\dagger} \, \mathcal{P}^{(1,2,3)} \right> \; \left< \tr U^{(2)} \, \mathcal{P}^{(2)} \right> \; \left< \tr U^{(3)} \, \mathcal{P}^{(3)} \right>  \point
\end{eqnarray}
Since the integration measure is invariant under the transformation $ U
\rightarrow U^\dagger $, the following equalities hold:
\begin{eqnarray}
&& \left< \tr U^{(1)\dagger} \, \tr U^{(2)\dagger} \, \tr U^{(3)\dagger} \, \mathcal{P}^{(1,2,3)} \right> = \left< \tr U^{(1)} \, \tr U^{(2)} \, \tr U^{(3)} \, \mathcal{P}^{(1,2,3)\dagger} \right> \nonumber \\
&& \mathcal{G}_{\textrm{Adj}} \simeq \frac{1}{8} iJ c_1 c_2 c_3 \left< \tr U^{(E)} \, \mathcal{P}^{(E,1)} \, \tr U^{(1)} \right> \nonumber \\
&& \qquad \left< \tr U^{(1)} \, \tr U^{(2)} \, \tr U^{(3)} \, \mathcal{P}^{(1,2,3)\dagger} \right> \; \left< \tr U^{(2)} \, \mathcal{P}^{(2)} \right> \; \left< \tr U^{(3)} \, \mathcal{P}^{(3)} \right>
\end{eqnarray}
the latter being equal to the planar limit of the following graph (represented in the right side of figure \ref{equivalence}) in AsQCD:
\begin{equation}
\mathcal{G}_{\textrm{As}} = c_1 c_2 c_3 \left< \tr U^{(E)} \, \mathcal{W}_{\textrm{As}}^{(1)} \, \mathcal{W}_{\textrm{As}}^{(2)} \, \mathcal{W}_{\textrm{As}}^{(3)} \, \mathcal{P}^{(E,1)} \, \mathcal{P}^{(1,2,3)\dagger} \, \mathcal{P}^{(2)} \, \mathcal{P}^{(3)} \right>_c \point
\end{equation}

This mechanism does not work with subleading graphs. It is enough to see the
graph in the figure \ref{loop}b to realize that no connected component can be
consistently reversed to get a graph of AsQCD.

This result has a general validity. Since the integration measure is invariant
under the substitution $ U \rightarrow U^\dagger $, one can reverse all the
directions in a connected component, without changing the value of the graph.
Since planar graphs do not contain loops, one can independently choose which
connected components to reverse. Reversing the directions of some components
is equivalent to interchanging $ \mathcal{W}_{\textbf{As}} \leftrightarrow
\frac{1}{2} \mathcal{W}_{\textbf{Adj}} $. In this way, one can change the
representation of the fermion and interchange Dirac with Majorana fermion. No
change in the coefficients $ c(\alpha) $ is needed, because they are
representation-independent. In conclusion, planar equivalence comes from the
graph-by-graph equality of expectation values of the two theories.

\section{Conclusions}

The results obtained are based on the possibility of expanding the expectation
values as power series in $\lambda^{-1}$ and $\kappa$. Of course, a phase
transition can exist in the plane $(\lambda,\kappa)$. Thus, the present work
proves the orientifold planar equivalence only in the phase containing the
point $\lambda=\infty$ and $\kappa=0$.

It is known that pure $SU(N)$ gauge theory in two dimensions on the lattice
has a phase transition in the planar limit (also at a finite volume) between a
strong-coupling and a weak-coupling phase \cite{Gross:1980he}. Moreover,
Kiskis, Narayanan and Neuberger \cite{Kiskis:2003rd,Narayanan:2005en} showed
numerically that such a phase transition exists in four dimensions, too.

At the moment, the phase structure of gauge theories with fermions in a
two-indices representation is not known.

Clearly, it is also conceivable that the strong-coupling and large-mass phase
does not contain the continuum limit (that is at $\lambda=0$ and
$\kappa=\infty$). In this case, no information can be inferred on the validity
of the orientifold planar equivalence between the two theories on the
continuum.

\end{document}